\documentclass[12pt]{article}
\usepackage[dvips]{graphicx}
\def\doublespace{\lineskip      .25 ex\baselineskip 3.0
ex\lineskiplimit 0 ex\parskip 1.0 ex plus.50 ex minus .25 ex}%

\begin{document}
\doublespace

\title{Non-uniform Braneworld Stars:\\ an Exact Solution }
\author{ J.
Ovalle\footnote{jovalle@usb.ve; jovalle@fisica.ciens.ucv.ve}
 \\
\vspace*{.25cm}\\
 Escuela de F\'{\i}sica, Universidad Central de Venezuela, \\ Caracas,
 Venezuela.
 }
\date{}
\maketitle
\begin{abstract}

In this paper the first exact interior solution to Einstein's
field equations for a static and non-uniform braneworld star with
local and non-local bulk terms is presented. It is shown that the
bulk Weyl scalar ${\cal U}(r)$ is always negative inside the
stellar distribution, in consequence it reduces both the effective
density and the effective pressure. It is found that the
anisotropy generated by bulk gravity effect has an acceptable
physical behaviour inside the distribution. Using a
Reissner-N\"{o}rdstrom-like exterior solution, the effects of bulk
gravity on pressure and density are found through matching
conditions.
\end{abstract}
keywords: General Relativity; Braneworld; Exact Solutions.
\newpage
\section{Introduction}
\label{intro} \vspace{0.2cm}

The consequences of the braneworld theory \cite{rs} in general
relativity have been studied with great interest during the past
years \cite{chr} (see also a review paper \cite{maartRev2004} and
references therein). Its possible effects on our observable 4D
universe have been extensively studied through cosmological
scenarios \cite{cosmobrane} (see also \cite{cosmo04} and
references therein). The study of the consequences in astrophysics
\cite{astrobrane} has been mostly limited to the exterior region,
even though it is well known that gravitational collapse could
produce very high energies in the interior, where the braneworld
corrections to general relativity would become significant
\cite{astrobrane2} (see \cite{RuthGregory08} for a recent study of
black holes on the brane).

In the astrophysics context, the general fact that a single 5D
solution can generate different scenarios in 4D \cite{ponce05},
makes the analysis in the exterior region particularly attractive
in searching for solutions beyond Schwarzschild's. Hence several
scenarios which might be useful in predicting observable effects
from the extra dimension can be considered. On the other hand, the
study on internal stellar structure in the braneworld remains
unknown so far. The reason is very simple: the internal stellar
structure is more complex that any other scenario, thus the high
energy and nonlocal corrections to Einstein's field equation,
which in general lead to a complicated indefinite system of
equations in the brane, produce an even more complicated system of
equations in the interior. Hence the construction and eventual
study of internal stellar solutions in the brane is a difficult
process, except when uniform distributions are
considered\footnote{In the pioneer work of Germani and Maartens
\cite{germ} exact solutions for a uniform distribution were found,
where it can be seen how much more difficult it would be to find a
solution except for uniform stellar distributions.}. Consequently
there are some important questions which remain without answer.
For instance, the role played by density gradients as a source of
Weyl stresses, and its eventual consequence on the gravitational
collapse remain unknown. An internal consistent solution would be
useful in order to consider these issues. However, in general,
finding a consistent solution in the brane, no matter which
scenario is being considered, represents a challenge which final
answer require more information on the 5D geometry and the way on
how our 4D spacetime is embedded in the bulk.

There is a large number of studies on the consistency of
braneworld mo\-dels in the brane. For instance, the approach
developed in the series of papers by Mukohyama \cite{Mukohyama}
has been extensively used in calculating gravity perturbations in
a braneworld scenario. Early works on consistent linearized
gra\-vity \cite{AIMVV00}, tests of consistency beyond linear order
\cite{gren00}, stability and energy conservation \cite{Sasaki00},
thus as low energy effective theory in the context of the
braneworld \cite{kanno02}-\cite{sk03}, have been extensively used
as well. On the other hand, and from the point of view of a brane
observer, there is an issue which must be faced in the search of
any consistent solution in the brane; namely, the non locality and
non closure of the braneworld equations
\cite{SMS00}-\cite{koyamasoda}. This is an open problem for which
a solution requires a better understanding of the bulk geometry
and proper boundary conditions\footnote{For a discussion about the
role played by the boundary conditions and the gravitational
instability on the brane see \cite{shtanov1}}. It is well known
that the source of the non locality and non closure of the
braneworld equations is directly related with the projection
${\cal E}_{\mu\nu}$ of the bulk Weyl tensor on the brane. Several
ways have been taken to overcome this problem, most of them based
in restrictions on the tensor ${\cal E}_{\mu\nu}$. For instance a
restriction which has proven to be useful consists in discarding
the anisotropic stress associated to ${\cal E}_{\mu\nu}$
\cite{shtanov07}. A different and more radical restriction on
${\cal E}_{\mu\nu}$ consist in imposing the constraint ${\cal
E}_{\mu\nu}=0$. However this condition, which was initially used
in some papers, is incompatible with the Bianchi identity in the
brane \cite{koyama05}.

In this paper there will be no direct restrictions on ${\cal
E}_{\mu\nu}$. Instead of this approach, the method developed in
Ref.\cite{jovalle07} shall be used to solve the non closure
problem of braneworld equations. This method, based in the fact
that any stellar solution on the brane must have the general
relativity solution as a limit, is used in this paper to generate
the first exact and physically acceptable interior solution for a
nonuniform stellar distribution having local and non-local bulk
terms on the brane.

This paper is organized as follows. In Section \textbf{2} the
Einstein field equations and matching conditions in the brane for
a spherically symmetric distribution is reminded. In Section
\textbf{3} a regular and physically acceptable exact internal
solution is found. In Section \textbf{4} an analysis of the
solution is carried out. In the last section the conclusions are
presented.

\section{The field equations and matching conditions}
\label{field equations}

The Einstein field equations on the brane may be written as a
modification of the standard field equations
\cite{SMS00}$^{,}$\cite{maart2001}
\begin{equation}\label{einst}
G_{\mu\nu}=-8\pi T_{\mu\nu}^{T}-\Lambda g_{\mu\nu},
\end{equation}
where $\Lambda$ is the cosmological constant on the brane. The
energy-momentum tensor has new terms carrying bulk effects onto
the brane:
\begin{equation}\label{tot}
T_{\mu\nu}\rightarrow T_{\mu\nu}^{\;\;T}
=T_{\mu\nu}+\frac{6}{\sigma}S_{\mu\nu}+\frac{1}{8\pi}{\cal
E}_{\mu\nu},
\end{equation}
here $\sigma$ is the brane tension. The new terms $S_{\mu\nu}$ and
$\cal{E}_{\mu\nu}$ are the high-energy corrections and $KK$
corrections respectively, and are given by
\begin{equation}\label{s}
S_{\mu\nu}=\frac{1}{12}T_\alpha^{\;\alpha}T_{\mu\nu}-\frac{1}{4}T_{\mu\alpha}T^\alpha{\;\nu}
+\frac{1}{24}g_{\mu\nu}\left[3T_{\alpha\beta}T^{\alpha\beta}-(T_\alpha^{\;\alpha})^2\right],
\end{equation}
\begin{equation}\label{e}
-8\pi{\cal E}_{\mu\nu}=-\frac{6}{\sigma}\left[{\cal U}(u_\mu
u_\nu+\frac{1}{3}h_{\mu\nu})+{\cal P}_{\mu\nu}+{\cal
Q}_{(\mu}u_{\nu)}\right],
\end{equation}
being ${\cal U}$ the bulk Weyl scalar and ${\cal P}_{\mu\nu}$ and
${\cal Q}_\mu $ the anisotropic stress and energy flux
respectively.

We consider a spherically symmetric static distribution, hence
$Q_\mu =0$ and
\begin{equation}
{\cal P}_{\mu\nu}={\cal P}(r_\mu r_\nu+\frac{1}{3}h_{\mu\nu}),
\end{equation}
where $r_\mu$ is a unit radial vector and
$h_{\mu\nu}=g_{\mu\nu}-u_\mu u_\nu$ the projection tensor with
4-velocity $u^{\mu}$. The line element is given in
Schwarzschild-like coordinates by
\begin{equation}
\label{metric}ds^2=e^\nu dt^2-e^\lambda dr^2-r^2\left( d\theta
^2+\sin {}^2\theta d\phi ^2\right)
\end{equation}
where $\nu$ and $\lambda$ are functions of $r$.

The metric (\ref{metric}) has to satisfy (\ref{einst}). In our
case with $\Lambda=0$ we have:

\begin{equation}
\label{ec1}-8\pi \left( \rho
+\frac{1}{\sigma}\left(\frac{\rho^2}{2}+\frac{6}{k^4}\cal{U}\right)
\right) =-\frac 1{r^2}+e^{-\lambda }\left( \frac
1{r^2}-\frac{\lambda _1}r\right),
\end{equation}

\begin{equation}
\label{ec2}-8\pi
\left(-p-\frac{1}{\sigma}\left(\frac{\rho^2}{2}+\rho p
+\frac{2}{k^4}\cal{U}\right)-\frac{4}{k^4}\frac{\cal{P}}{\sigma}\right)
=-\frac 1{r^2}+e^{-\lambda }\left( \frac 1{r^2}+\frac{\nu
_1}r\right),
\end{equation}

\begin{eqnarray}
\label{ec3}-8\pi
\left(-p-\frac{1}{\sigma}\left(\frac{\rho^2}{2}+\rho p
+\frac{2}{k^4}\cal{U}\right)+\frac{2}{k^4}\frac{\cal{P}}{\sigma}\right)
= \frac 14e^{-\lambda }\left[ 2\nu _{11}+\nu _1^2-\lambda _1\nu
_1+2 \frac{\left( \nu _1-\lambda _1\right) }r\right],
\end{eqnarray}

\begin{equation}
\label{con1}p_{1}=-\frac{\nu_1}{2}(\rho+p),
\end{equation}
where $f_1\equiv df/dr$ and $k^2=8{\pi}$. The general relativity
is regained when $\sigma^{-1}\rightarrow 0$ and (\ref{con1})
becomes a linear combination of (\ref{ec1})-(\ref{ec3}).

The Israel-Darmois matching conditions at the stellar surface
$\Sigma$ give
\begin{equation}
\label{matching1} [G_{\mu\nu}r^\nu]_{\Sigma}=0
\end{equation}
where $[f]_{\Sigma}\equiv f(r)\mid_{R^+}-f(r)\mid_{R^-}$ Using
(\ref{matching1}) and the field equation (\ref{einst}) with
$\Lambda=0$ we have
\begin{equation}
\label{matching2} [T^{\;\;T}_{\mu\nu}r^\nu]_{\Sigma}=0,
\end{equation}
which leads to
\begin{equation}
\label{matching3} \left[
\left(p+\frac{1}{\sigma}\left(\frac{\rho^2}{2}+\rho p
+\frac{2}{k^4}\cal{U}\right)+\frac{4}{k^4}\frac{\cal{P}}{\sigma}\right)
\right]_{\Sigma}=0.
\end{equation}
This takes the final form
\begin{equation}
\label{matchingf}
p_R+\frac{1}{\sigma}\left(\frac{\rho_R^2}{2}+\rho_R p_R
+\frac{2}{k^4}{\cal U}_R^-\right)+\frac{4}{k^4}\frac{{\cal
P}_R^-}{\sigma} = \frac{2}{k^4}\frac{{\cal
U}_R^+}{\sigma}+\frac{4}{k^4}\frac{{\cal P}_R^+}{\sigma},
\end{equation}
where $f_R\equiv f(r)\mid_{r=R}$. The equation (\ref{matchingf})
gives the general matching condition for any static spherical
braneworld star\footnote{The general matching conditions on the
brane for a spherically symmetric vacuum region embedded into a
cosmological environment can be seen in \cite{gergely2005}}
\cite{germ} . When $\sigma^{-1}\rightarrow 0$ we obtain the well
known matching condition $p_R =0$. In the particular case of the
Schwarzschild exterior solution ${\cal U}^+={\cal P}^+ =0$, the
matching condition (\ref{matchingf}) becomes:
\begin{equation}
\label{matchingfS}
p_R+\frac{1}{\sigma}\left(\frac{\rho_R^2}{2}+\rho_R p_R
+\frac{2}{k^4}{\cal U}_R^-\right)+\frac{4}{k^4}\frac{{\cal
P}_R^-}{\sigma} = 0.
\end{equation}
Thus the matching conditions do not have a unique solution on the
brane.

It is easily seen that the field equations (\ref{ec1})-(\ref{ec3})
can be written as
\begin{equation}
\label{usual} e^{-\lambda}=1-\frac{8\pi}{r}\int_0^r r^2\left[\rho
+\frac{1}{\sigma}\left(\frac{\rho^2}{2}+\frac{6}{k^4}\cal{U}\right)\right]dr,
\end{equation}
\begin{equation}
\label{pp}\frac{8\pi}{k^4}\frac{{\cal
P}}{\sigma}=\frac{1}{6}\left(G_1^1-G_2^2\right),
\end{equation}
\begin{equation}
\label{uu}\frac{6}{k^4}\frac{{\cal
U}}{\sigma}=-\frac{3}{\sigma}\left(\frac{\rho^2}{2}+\rho
p\right)+\frac{1}{8\pi}\left(2G_2^2+G_1^1\right)-3p
\end{equation}
with
\begin{equation}
\label{g11} G_1^1=-\frac 1{r^2}+e^{-\lambda }\left( \frac
1{r^2}+\frac{\nu _1}r\right),
\end{equation}
\begin{equation}
\label{g22} G_2^2=\frac 14e^{-\lambda }\left[ 2\nu _{11}+\nu
_1^2-\lambda _1\nu _1+2 \frac{\left( \nu _1-\lambda _1\right)
}r\right].
\end{equation}
The equation (\ref{usual}) actually represents an integral
differential equation for the geometrical function $\lambda(r)$,
something completely different from the general relativistic case,
and a direct consequence of the non locality of the braneworld
equations. The only solution known for this equation is given as
\cite{jovalle07}
\begin{equation}
\label{edlrwss} e^{-\lambda}={1-\frac{8\pi}{r}\int_0^r r^2\rho
dr}+e^{-I}\int_0^r\frac{e^I}{(\frac{\nu_1}{2}+\frac{2}{r})}\left[H(p,\rho,\nu)+\frac{8\pi
}{\sigma}\left(\rho^2+3\rho p\right)\right]dr,
\end{equation}
with
\begin{equation}
\label{finalsol} H(p,\rho,\nu)\equiv 8\pi
3p-\left[\mu_1(\frac{\nu_1}{2}+\frac{1}{r})+\mu(\nu_{11}+\frac{\nu_1^2}{2}+\frac{2\nu_1}{r}+\frac{1}{r^2})-\frac{1}{r^2}\right],
\end{equation}
where
\begin{eqnarray} \label{I} I\equiv
\int\frac{(\nu_{11}+\frac{\nu_1^2}{2}+\frac{2\nu_1}{r}+\frac{2}{r^2})}{(\frac{\nu_1}{2}+\frac{2}{r})}dr,
\end{eqnarray}
and
\begin{equation}
\mu\equiv 1-\frac{8\pi}{r}\int_0^r r^2\rho dr.
\end{equation}
The function $H(p,\rho,\nu)$ measures the anisotropic effects due
to bulk consequences on $p$, $\rho$ and $\nu$, and its physical
meaning will be useful in the searching of exact solutions. This
will be addressed in the next section.

\section{An exact solution}

So far we have the interior Weyl functions ${\cal P}$ and ${\cal
U}$ plus the geometric function $\lambda(r)$, respectively given
by (\ref{pp}), (\ref{uu}) and (\ref{edlrwss}), and three unknown
functions $\{p(r),\rho(r),\nu(r)\}$ satisfying one equation,
namely, the conservation equation (\ref{con1}). Therefore it is
necessary to prescribe additional information to close the system.
First of all, to ensure the correct limit at low energies, the
following constraint is imposed on the brane
\begin{equation}
\label{constraint}
 H(p,\rho,\nu)=0.
\end{equation}
The constraint (\ref{constraint})  has been proven to be useful in
finding solutions which possesses general relativity as a limit
\cite{jovalle07}, and has a clear physical interpretation:
eventual bulk consequences on $p$, $\rho$ and $\nu$ do not produce
anisotropic effects on the brane. This constraint ensures that a
braneworld solution be consistent with general relativity, as is
shown in Ref.\cite{jovalle07}. Unfortunately this constraint is
not enough, and one additional condition must be imposed.
Therefore, from the point of view of a brane observer, many
solutions are possible. However not all of these solutions are of
physical interest. Hence the brane observer has to impose a
condition in the brane which must lead to a physically acceptable
solution, namely, regular at the origin, pressure and density
defined positive, well defined mass and radius, monotonic decrease
of the density and pressure with increasing radius, etc. All these
conditions reduce enormously the possible solutions $(p,\rho,\nu)$
to $(\ref{con1})$ and $(\ref{constraint})$, even more in the
searching of an exact solution.

It might be logical to think of the Schwarzschild condition
$e^{\nu}=e^{-\lambda}$ to close the system. However this would
produce a very complicated integral differential equation for
$\lambda$, as can be seen through (\ref{edlrwss}). On the other
hand, it is clear that to find an exact expression for the
geometric function $\lambda(r)$, given by (\ref{edlrwss}), an
analytic solution for (\ref{I}) is needed. Keeping this in mind, a
huge and simple family of exact solutions for (\ref{I}) is
considered, given by
\begin{equation}
\label{regularmet00} e^{\nu}=A(1+Cr^{m})^n,
\end{equation}
which is characterized by the constants $A$, $C$, $m$ and $n$. The
constants $A$ and $C$ are eventually found through matching
conditions, whereas $m$ and $n$ are parameters to be used in the
searching of an exact solution for $p(r)$ and $\rho(r)$.

Using (\ref{regularmet00}) in (\ref{con1}) the pressure is found
in terms of the density
\begin{equation}\label{prev2}
p(r)=\frac{ 2\,B -
      {\sqrt{A}}\,C\,m\,n\,
       \int r^{m-1}\,
           {\left( 1 + C\,r^m \right) }^
            {\frac{n}{2}-1}\,\rho(r)\,d
          r  }{2\,{\sqrt{A}}\,
    {\left( 1 + C\,r^m \right) }^
     {\frac{n}{2}}},
\end{equation}
with $B$ a constant of integration.

Now using (\ref{regularmet00}) and (\ref{prev2}) in the constraint
(\ref{constraint}), the following integral equation for the
density is obtained
\begin{eqnarray}\label{intdiffeq} && - \frac{
     \left( r - 8\,\pi \,
        \int_0^r r^2\,\rho(r)\,dr \right) }
     {2\,r^3\,{\left( 1 + C\,r^m \right)
         }^2}\left[2 +
       2\,C\,\left( 2 + m\,n + m^2\,n
          \right) \,r^m \right.
\nonumber \\
 && \left.
       +C^2\,\left( 2 + 2\,m\,n +
          m^2\,n^2 \right) \,r^{2\,m}\right]
\nonumber \\
         &&- \frac{12\,\pi \,
     \left( -2\,B +
       {\sqrt{A}}\,C\,m\,n\,
        \int_0^r r^{-1 + m}\,
            {\left( 1 + C\,r^m \right) }^
             {-1 + \frac{n}{2}}\,\rho(r)\,
           dr \right) }{{\sqrt{A}}\,
     {\left( 1 + C\,r^m \right) }^
      {\frac{n}{2}}}
\nonumber \\
&&-\frac{4\,\pi \,
     \left( 2 + C\,
        \left( 2 + m\,n \right) \,r^m
       \right) \,
     \left( \int_0^r r^2\,\rho(r)\,dr -
       r^3\,\rho(r) \right) }{r^3\,
     \left( 1 + C\,r^m \right) }+r^{-2}=0.
\end{eqnarray}
Hence $\rho(r)$ can be found by (\ref{intdiffeq}) and then $p(r)$
through (\ref{prev2}). In this stage it is important to stress
that the goal of this paper is to find a simple exact braneworld
solution, hence we are not interested in the most general solution
$\rho(r)$ to the integral equation (\ref{intdiffeq}). Indeed, a
particularly useful solution for (\ref{prev2}) and
(\ref{intdiffeq}), leading to an exact braneworld solution through
(\ref{edlrwss}), will be constructed. This is shown below.

The first step is to examine the complicated bulk contribution to
$\lambda(r)$ in (\ref{edlrwss}). When the constraint
(\ref{constraint}) is imposed to ensure general relativity as a
limit, the geometric function $\lambda(r)$ underwent a great
simplification, leaving only high energy corrective terms. Then
when (\ref{regularmet00}) is used, the complicated integral
expression in (\ref{edlrwss}) is reduced even more, leading to
\begin{eqnarray}
\label{edlrwss2}
\int_0^r\frac{e^I}{(\frac{\nu_1}{2}+\frac{2}{r})}\left(\rho^2+3\rho
p\right)dr=&&\int_0^r2r^2\,{\left( 1 + C\,r^m \right) }^
   {n-1}{\left[ 4 +
      C\left( 4 + m\,n \right)r^m
      \right] }^
   {\frac{4 + n\left( m-3 \right) }
     {4 + m\,n}}
\nonumber \\ &&\left(\rho^2+3\rho p\right)dr.
\end{eqnarray}
Finding a physically acceptable exact solution for (\ref{prev2})
and (\ref{intdiffeq}), which can produce an exact solution for
(\ref{edlrwss2}), represents a difficult process. Hence the
integral expression (\ref{edlrwss2}) has to be simplified as much
as possible. Indeed,  the integral (\ref{edlrwss2}) is
tremendously simplified when $4 + n\left( m-3 \right)=0$, leading
to
\begin{eqnarray}
\label{edlrwss3}
\int_0^r\frac{e^I}{(\frac{\nu_1}{2}+\frac{2}{r})}\left(\rho^2+3\rho
p\right)dr=&&\int_0^r2r^2\,{\left( 1 + C\,r^m \right) }^
   {\frac{1+m}{3-m}}\left(\rho^2+3\rho p\right)dr.
\end{eqnarray}
The next step will be to consider in (\ref{prev2}) a simple ansatz
for $\rho(r)$ which is capable of producing a simple expression
for the pressure, and then to use (\ref{intdiffeq}) to fix the
parameters of the ansatz. The idea is to obtain a solution for
both $p(r)$ and $\rho(r)$ as simple as possible such that the
integral (\ref{edlrwss3}) has an analytic expression. If this is
accomplished, the geometric function $\lambda(r)$, given through
(\ref{edlrwss}), and both interior Weyl functions ${\cal P}$ and
${\cal U}$, given by (\ref{pp}) and (\ref{uu}) respectively, will
have exact expressions.

By simple inspection of (\ref{prev2}) it is easy to see that a
convenient ansatz for $\rho(r)$ can be written by
\begin{equation}
\label{ansrho} \rho(r)=(1+Cr^m)^{-(n/2 + 1)}\sum_{s=0} a_s r^s.
\end{equation}
On the other hand, since (\ref{regularmet00}) must be regular at
the origin $r=0$, $m$ has to be positive. The constant $n$, which
is given by $n=\frac{4}{3-m}$, is considered positive to obtain a
pressure with a physically acceptable behaviour. Thus in our case
$m$ satisfies $0<m<3$. Taking $m=2$ and using (\ref{ansrho}) in
(\ref{prev2}), it is found that $p(r)$ has a relatively simple
expression free of special functions when \footnote{Using an
algebraic manipulator, it is not complicated to realize that
keeping a generic value for $m$ and $s$ produce a solution to
(\ref{prev2}) having special functions. This special functions
eventually would make impossible an exact solution to $\lambda(r)$
through (\ref{edlrwss}).}
\begin{equation}
\label{ansrho2} \rho(r)=\frac{(a_0+a_2r^2+a_4r^4)}{(1+Cr^2)^3}.
\end{equation}
Hence the following expression for the pressure is found
\begin{equation}
\label{prev22} p(r)=\frac{C_0
 +C_1\,r^2+C_2\,r^4 +
    2\,\left( 2\,a_4 -
       a_2\,C \right) \,
     \left( 1 + C\,r^2 \right) \,
     \log (1 + C\,r^2)}{\,C^2\,
    {\left( 1 + C\,r^2 \right) }^3},
\end{equation}
where
\begin{eqnarray}
\label{constccc} C_0&=&2 a_4 -
       a_2\,C +
       a_0\,C^2 +
       \frac{B\,C^2}{{\sqrt{A}}};\;\; C_1=\frac{B\,C^3}{{\sqrt{A}}}- 2\,a_4\,C,
\nonumber \\
C_2&=& -
    2\,a_4\,C^2.
\end{eqnarray}
The expression (\ref{prev22}) will hardly produce an exact
function to (\ref{edlrwss3}) unless the logarithmic function be
removed. Thus,
\begin{equation}
\label{nolog}  a_4=\frac{C}{2}a_2.
\end{equation}
Now using (\ref{ansrho2}), the constraint (\ref{intdiffeq}) is
written as
\begin{eqnarray}
\label{intdiff3} &&{\sqrt{C}}\,\left( -24\,C^2 +
     66\,a_2\,\pi  +
     36\,a_0\,C\,\pi  +
     \frac{24\,B\,C\,\pi }{{\sqrt{A}}}
     \right) \,r +
  {\sqrt{C}}\,\left( -20\,C^3 +
     70\,a_2\,C\,\pi  \right)r^3
\nonumber \\ &&
     - \left( 9\,a_2 -
     2\,a_0\,C \right) \,\pi \,
   \left( 5 + 9\,C\,r^2 \right) \,
   \arctan ({\sqrt{C}}\,r)=0,
\end{eqnarray}
hence it is found that
\begin{equation}
\label{constv2}
a_0=\frac{9\,C}{7\,\pi};\;\;\;a_2=\frac{2\,C^2}{7\,\pi};\;\;\;a_4=\frac{C^3}{7\,\pi};\;\;\;B=-\frac{12\,C\,\sqrt{A}}{7\,\pi}.
\end{equation}

Using (\ref{constccc}) and (\ref{constv2}) in (\ref{ansrho2}) and
(\ref{prev22}), a simple and physically acceptable expression is
found for both the pressure and density. Thus the solution for
(\ref{con1}) and (\ref{constraint}) is finally written as
\footnote{The expression for $\nu$ and $\rho$ were found in
\cite{DF85} for a perfect fluid in the context of general
relativity. However the solution shown there is not physically
acceptable due to $G_r^r\neq G_\theta^\theta$.}
\begin{equation}
\label{regularmet000} e^{\nu}=A(1+Cr^{2})^4
\end{equation},
\begin{equation}\label{regularden}
\rho(r)=\frac{C\,\left( 9 + 2\,C\,r^2 +
      C^2\,r^4 \right) }{7\,\pi \,
    {\left( 1 + C\,r^2 \right) }^3}
\end{equation}
and
\begin{equation}\label{regularpress}
p(r)=\frac{2C(2-7Cr^2-C^2r^4)}{7\pi(1+Cr^2)^3},
\end{equation}
leaving $A$ and $C$ to be determined by matching conditions.

Using (\ref{regularmet000})-(\ref{regularpress}) in
(\ref{edlrwss}) a regular and well defined solution for
$\lambda(r)$ is obtained
\begin{equation}\label{reglambda}
e^{-\lambda(r)}=1-\frac{2\tilde{m}(r)}{r},
\end{equation}
where the interior mass function $\tilde{m}$ is given by
\begin{eqnarray}\label{regularmass}
\tilde{m}(r)&=&m(r)-\frac{1}{\sigma}\left(\frac{2}{7}\right)^2\frac{Cr}{2\pi}\left[\frac{240+589Cr^2-25C^2r^4-41C^3r^6-3C^4r^8}{3(1+Cr^2)^4(1+3Cr^2)}\right.
\nonumber \\&&
\left.-\frac{80}{(1+Cr^2)^2}\frac{arctg(\sqrt{C}r)}{(1+3Cr^2)\sqrt{C}r}\right],
\end{eqnarray}
with $m(r)$ being the general relativity interior mass function,
given by the standard form
\begin{equation}
\label{regularmass2} m(r)=\int_0^r 4\pi
r^2{\rho}dr=\frac{4}{7}Cr^3\frac{(3+Cr^2)}{(1+Cr^2)^2},
\end{equation}
hence the total general relativity mass is obtained
\begin{equation}
\label{regtotmass} M\equiv
m(r)\mid_{r=R}=\frac{4}{7}CR^3\frac{(3+CR^2)}{(1+CR^2)^2},
\end{equation}
where $R$ is the radius of the distribution.

Using (\ref{pp}) and (\ref{uu}) a regular solution for the
interior Weyl functions is obtained
\begin{eqnarray}
\label{regP} {\cal P}(r)&=&\frac{32}{441
r^3(1+Cr^2)^6(1+3Cr^2)^2}\left[Cr\left(180+2040Cr^2+8696C^2r^4\right.\right.\nonumber\\&&
\left.\left.+16533C^3r^6+12660C^4r^8+146C^5r^{10}-120C^6r^{12}+9C^7r^{14}\right)\right.
\nonumber \\ &&
\left.-60\sqrt{C}(1+Cr^2)^3(3+26Cr^2+63C^2r^4)arctg(\sqrt{C}r)\right],
\end{eqnarray}

\begin{eqnarray}\label{regU}
{\cal U}(r)&=&\frac{32}{441
r(1+Cr^2)^6(1+3Cr^2)^2}\left[C^2r\left(795+4865Cr^2+10044C^2r^4\right.\right.
\nonumber \\ &&
\left.\left.+6186C^3r^6-373C^4r^8-219C^5r^{10}-18C^6r^{12}\right)\right.
\nonumber \\ &&
\left.-240C^{3/2}(1+Cr^2)^3(5+9Cr^2)arctg(\sqrt{C}r)\right].
\end{eqnarray}

The expressions (\ref{regularmet000})-(\ref{regularpress}) with
(\ref{regP}) and (\ref{regU}) represent an {\it exact analytic
solution} to the system (\ref{ec1})-(\ref{con1}).

\section{Analysis of the solution}

As can be seen by figure 1, the scalar function ${\cal U}(r)$ is
always negative inside the stellar distribution, with a maximum
negative value at the origin $r=0$. This situa\-tion may be
explained through the general expression for ${\cal U}(r)$, given
by Eq. (\ref{uu}). It shows two "sources" for ${\cal U}(r)$: the
first kind given by the first two terms in the second hand side of
Eq. (\ref{uu}), which are high energy corrections, always
negative. The second kind is given by the remaining terms, which
clearly represent an anisotropic expression, which is always
positive. Hence when the anisotropy projected onto the brane is
not high enough, the dominant high energy terms produce a negative
scalar function ${\cal U}(r)$, which is the case presented here.
This negative scalar function reduces both the effective density
and the effective pressure, as can be seen by the field equations
(\ref{ec1})-(\ref{ec3}). On the other hand, the anisotropy inside
the braneworld star is shown by the figure 2. It increases until
reaches a maximum value, then decreases until ${\cal P}=0$ at
$r=0$. This is directly connected with the correction for
$\lambda$ proportional to high energy terms shown in
(\ref{edlrwss}). This correction is the only bulk effect underwent
by the metric when the constraint (\ref{constraint}) is imposed,
therefore it represents the only source for ${\cal P}$, as can be
clearly seen through (\ref{pp}).

The bulk contribution to $p$, $\rho$ and $\nu$ is found by
matching conditions, where the assumption of vanishing pressure at
the surface will be dropped \cite{deru}$^{,}$\cite{gergely2007}.
As Schwarzschild is not the only possible static exterior
solution, we have many scenarios to consider. For instance let us
consider the Schwarzschild exterior solution
\begin{equation}
\label{SchwExt}
e^{\nu^{+}}=e^{-\lambda^{+}}=1-\frac{2\cal{M}}{r};\;\;\;\;\;\;\;{\cal
U}^{+}={\cal P}^{+}=0.
\end{equation}
The matching condition $[ds^2]_{\Sigma}=0$ at the stellar surface
$\Sigma$ yields
\begin{equation}
\label{RegmatchSchw1} A=(1-\frac{2\cal{M}}{R})(1+CR^{2})^{-4},
\end{equation}

\begin{eqnarray}\label{RegmatchSchw2}
\frac{2\cal{M}}{R}&=&\frac{2M}{R}-\frac{1}{\sigma}\left(\frac{2}{7}\right)^2\frac{C}{\pi}\left[\frac{240+589CR^2-25C^2R^4-41C^3R^6-3C^4R^8}{3(1+CR^2)^4(1+3CR^2)}\right.
\nonumber \\&&
\left.-\frac{80}{(1+CR^2)^2}\frac{arctg(\sqrt{C}R)}{(1+3CR^2)\sqrt{C}R}\right],
\end{eqnarray}
and using (\ref{matchingfS}) it is found that $C$ must satisfy the
condition
\begin{eqnarray}
\label{RegmatchSchw3}
 &\pi
 R\left[168CR^2+252(CR^2)^2-1848(CR^2)^3-4032(CR^2)^4-2352(CR^2)^5-252(CR^2)^6\right]\nonumber\\&
 +\frac{1}{\sigma}CR\left[240+2749CR^2+5276(CR^2)^2-266(CR^2)^3-372(CR^2)^4-27(CR^2)^5\right]\nonumber\\&
-\frac{1}{\sigma}240\sqrt{C}arctg(\sqrt{C}R)\left(1+11CR^2+19C^2R^4+9C^3R^6\right)=0,
\end{eqnarray}
hence solving (\ref{RegmatchSchw3}) $C$ is found as a function of
the brane tension $\sigma$.

In order to find the bulk contribution to $p$ and $\rho$ we need
to find $C$ satisfying (\ref{RegmatchSchw3}). To carry out this
the following solution is proposed
\begin{equation}
\label{RegC} C=C_0+\delta,
\end{equation}
where $C_0$ is the general relativity value of $C$, given by
\begin{equation}
\label{RegC0} C_0=\frac{\sqrt{57}-7}{2R^2},
\end{equation}
which is found using the condition $p(R)=0$ in
(\ref{regularpress}). In this sense $\delta$ represents the " bulk
perturbation" of the general relativity value of $C$. Using
(\ref{RegC}) in (\ref{RegmatchSchw3}) we have at first order in
$\sigma^{-1}$
\begin{equation}
\delta=\frac{-4\left[ \left( -236357 +
         31281\,{\sqrt{57}} \right) \,R -
      120\,\left( -9235 +
         1223\,{\sqrt{57}} \right) \,
       \arctan ({\sqrt{\frac{-7 +
              {\sqrt{57}}}{2}}}) \right] }
    {3 \pi\sigma R^5\left( -1261105 +
      167083\,{\sqrt{57}} \right) }
\end{equation}

The pressure can thus be found expanding $p(C)$ around $C_0$
\begin{equation}
\label{ExpanP}
p(C_0+\delta)=p(C_0)+\delta{\frac{dp}{dC}}\mid_{C=C_0},
\end{equation}
which leads to
\begin{equation}
\label{RegPSchw}
p(r)=\frac{2C_0}{7\pi}\frac{(2-7C_0r^2-C_0^2r^4)}{(1+C_0r^2)^3}+\frac{4}{7\pi}\frac{(1-9C_0r^2+2C_0^2r^4)}{(1+C_0r^2)^4}\delta.
\end{equation}
By the same way the density is found to be
\begin{equation}
\label{RegDSchw}
\rho(r)=\frac{C_0}{7\pi}\frac{(9+2C_0r^2+C_0^2r^4)}{(1+C_0r^2)^3}+\frac{1}{7\pi}\frac{(9-14C_0r^2+C_0^2r^4)}{(1+C_0r^2)^4}\delta.
\end{equation}
However at the surface and for any arbitrary $R$ always we have
\begin{equation}
\label{RegPSchw2}
p(R)=\frac{4}{7\pi}\frac{(1-9C_0R^2+2C_0^2R^4)}{(1+C_0R^2)^4}\delta
< 0.
\end{equation}
Hence the Schwarzschild exterior solution is incompatible with the
interior solution found here. Thus a different exterior solution
must be considered.

Using now the Reissner-N\"{o}rdstrom-like solution given in
\cite{dmpr}
\begin{equation}
\label{RegRNmet}
e^{\nu^+}=e^{-\lambda^+}=1-\frac{2\cal{M}}{r}+\frac{q}{r^2},
\end{equation}
\begin{equation}
\label{RegRNmet2} {\cal U}^+=-\frac{{\cal P}^+}{2}=\frac{4}{3}\pi
q\sigma\frac{1}{r^4},
\end{equation}
and considering the matching condition $[ds^2]_{\Sigma}=0$ at the
stellar surface $\Sigma$, we have
\begin{equation}
\label{RegmatchNR1}
A(1+CR^{2})^{4}=1-\frac{2\cal{M}}{R}+\frac{q}{R^2},
\end{equation}
\begin{eqnarray}\label{RegmatchNR2}
\frac{2\cal{M}}{R}&=&\frac{2M}{R}-\frac{1}{\sigma}\left(\frac{2}{7}\right)^2\frac{C}{\pi}\left[\frac{240+589CR^2-25C^2R^4-41C^3R^6-3C^4R^8}{3(1+CR^2)^4(1+3CR^2)}\right.
\nonumber \\&&
\left.-\frac{80}{(1+CR^2)^2}\frac{arctg(\sqrt{C}R)}{(1+3CR^2)\sqrt{C}R}\right]+\frac{q}{R^2},
\end{eqnarray}
and using (\ref{matchingf}) we obtain
\begin{eqnarray}
\label{qNR}
q&=&\frac{-4R}{147(1+CR^2)^5(1+3CR^2)}\left[CR\left((-2+CR^2+22C^2R^4+3C^3R^6)\right.\right.\nonumber
\\ &&
\left.84\pi
R^2(1+CR^2)^2+\frac{1}{\sigma}(-240-2749CR^2-5276C^2R^4+266C^3R^6\right.\nonumber
\\ &&
\left.\left.+372C^4R^8+27C^5R^{10})\right)+\frac{1}{\sigma}240\sqrt{C}(1+CR^2)^2(1+9CR^2)arctg(\sqrt{C}R)\right].
\nonumber
\\
\end{eqnarray}
The constants $\cal{M}$ and $q$ are given in terms of $C$ through
equations (\ref{RegmatchNR2}) and (\ref{qNR}) respectively, and
$C$ may be determined by (\ref{RegmatchNR1}) if $A$ is kept as a
free parameter, which can be used to find a physically acceptable
model. However we have to be aware of the fact that $A$ has a well
defined general relativity value, named $A_0$, which is given by
(\ref{RegmatchNR1}) at $\sigma^{-1}=0$
\begin{equation}
\label{RegmatchNR1GR} A_0(1+C_0R^{2})^{4}=1-\frac{2M}{R}.
\end{equation}
In this sense the free parameter associate to $A$, which will be
used to obtain an acceptable model, is the "bulk perturbation" of
$A$ given through
\begin{equation}
\label{Asolreg} A=A_0+\varepsilon.
\end{equation}

Using (\ref{RegC}) and (\ref{Asolreg}) in (\ref{RegmatchNR1}) we
obtain
\begin{equation}
\label{RegmatchNR1Pert}
(A_0+\varepsilon)[1+(C_0+\delta)R^{2}]^{4}=1-\frac{2{\cal
M}}{R}+\frac{q}{R^2}.
\end{equation}
Evaluating the expressions (\ref{RegmatchNR2}) and (\ref{qNR}) at
$C=C_0 +\delta$ and keeping li\-neal terms in $\sigma^{-1}$, the
equation (\ref{RegmatchNR1Pert}) leads to the explicit form to the
bulk perturbation of $C$, which is written as
\begin{equation}
\label{dNR}
\delta(\sigma)=\frac{7}{4}\frac{(1+C_0R^2)^3[\alpha(\sigma)-(1+C_0R^2)^4\varepsilon]}{(7+2C_0R^2+C_0^2R^4-2C_0^3R^6)},
\end{equation}
with
\begin{eqnarray}\label{alphaRN}
\alpha(\sigma)
&=&\frac{1}{\sigma}\left(\frac{2}{7}\right)^2\frac{C_0}{\pi}\left[\frac{240+589C_0R^2-25C_0^2R^4-41C_0^3R^6-3C_0^4R^8}{3(1+C_0R^2)^4(1+3C_0R^2)}\right.
\nonumber \\&&
\left.-\frac{80}{(1+C_0R^2)^2}\frac{arctg(\sqrt{C_0}R)}{(1+3C_0R^2)\sqrt{C_0}R}\right].
\end{eqnarray}
Hence giving by hand the perturbation $\varepsilon$ underwent by
$A$ due to the extra dimension, it is possible to obtain $\delta$
and thus the bulk consequences on $p$ and $\rho$ through
(\ref{RegPSchw}) and (\ref{RegDSchw}). The figure 3 shows the
behaviour of the pressure in both the general relativity and
braneworld case. It can be seen that the bulk gravity effect
reduces the pressure deep inside the distribution, but the
situation changes for the exterior layers, where the matching
conditions lead to $p\neq 0$ at the surface.

\begin{figure}
  \includegraphics{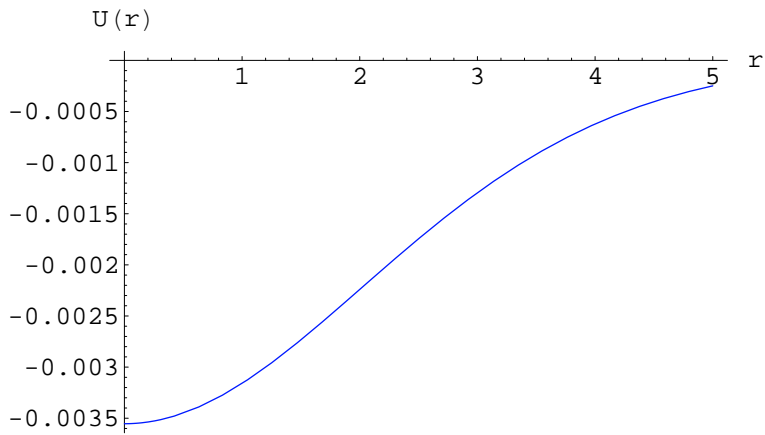}
\caption{The scalar function ${\cal U}(r)$ for a distribution with
$R=5$. ${\cal U}(r)$ is always negative in the interior, hence it
reduces both the effective density and effective pressure.}
\label{fig:1}       
\end{figure}

\begin{figure}
  \includegraphics{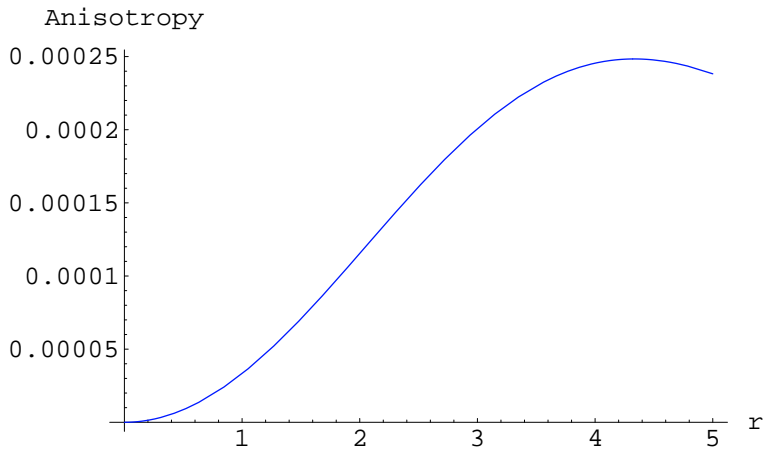}
\caption{Behaviour of the anisotropy ${\cal P}(r)$ inside the
stellar distribution with $R=5$. }
\label{fig:1}       
\end{figure}

\begin{figure}
  \includegraphics{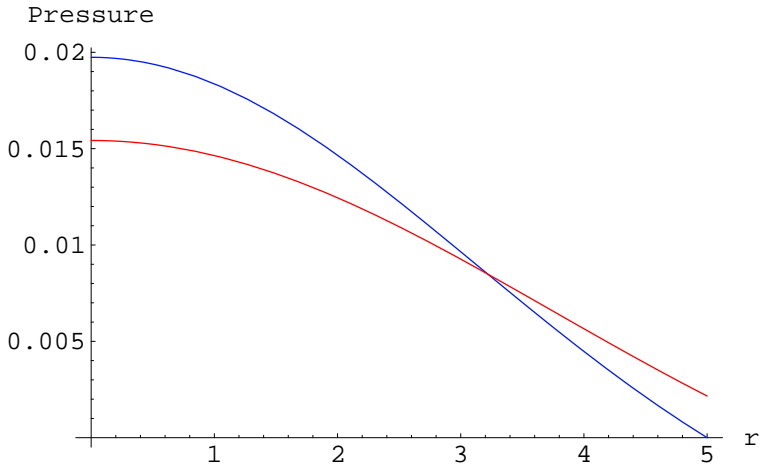}
\caption{Qualitative comparison of the pressure $p(r)$ in general
relativity
        ( $p(R)=0$ ) and the braneworld model ( $p(R)\neq 0$ ) with $R=5$.}
\label{fig:1}       
\end{figure}

\section{Conclusions}
In the context of the braneworld, a spherically symmetric, static
and non-uniform stellar distribution with Weyl stresses was
studied. The method developed in Ref. \cite{jovalle07}, which is
based in the fact that any stellar solution on the brane must have
the general relativity solution as a limit, was used to overcome
the non locality and non closure of the braneworld equations.
Hence there was no direct restriction on the projected ${\cal
E}_{\mu\nu}$ Weyl tensor on the brane, which is a method usually
used.

By prescribing the temporal metric component $g_{00}$, the first
exact and physically acceptable interior solution to Einstein's
field equations for a static and non-uniform braneworld star was
found. It was shown that this solution is incompatible with the
Schwarzschild's exterior metric. Using the
Reissner-N\"{o}rdstrom-like solution given in Ref. \cite{dmpr},
the effects of bulk gravity on pressure and density were found
through matching conditions, where the assumption of vanishing
pressure at the stellar surface was dropped. It was found that the
bulk gravity effect reduces the pressure deep inside the
distribution, but the situation changes for the exterior layers as
a direct consequence of matching conditions, in agreement with the
previous study in Ref. \cite{jovalle07}.

It was found that the Weyl scalar function ${\cal U}(r)$ is always
negative inside the stellar distribution. In consequence it
reduces both the effective density and the effective pressure. On
the other hand, the anisotropy inside the braneworld star, which
is directly connected with the deformation underwent by $\lambda$
due to bulk gravity effects, has an acceptable physical behaviour.

The exact solution found in this paper was possible as a direct
consequence of the constraint $H(p,\rho,\nu)= 0$. This essentially
allows us to simplify the solution for the geometric function
$\lambda(r)$ by eliminating some anisotropic effects on the brane.
As can be seen by Eq. (\ref{pp}), the source of ${\cal P}$ is the
deformation underwent by the geometric functions $\lambda(r)$ and
$\nu(r)$ due to bulk consequences on the brane. However when the
constraint $H(p,\rho,\nu)=0$ is imposed, the bulk effect on
$\nu(r)$ does not produce any anisotropic consequence, leaving
thus the deformation underwent by $\lambda(r)$ as the only source
of anisotropy on the brane. Furthermore, this constraint reduces
the deformation of $\lambda(r)$, leaving a corrective term
proportional to high-energy effects of bulk gravity, as can be
clearly seen through the equation (\ref{edlrwss}). Since the
constraint $H(p,\rho,\nu)=0$ removes all possible sources of
anisotropy except for the corrective term of $\lambda(r)$
proportional to high-energy effects, it follows that the
anisotropic effect of the bulk on the brane is reduced to its
minimal expression, hence the constraint imposed represents a {\it
condition of minimal anisotropy} on the brane.

The condition of minimal anisotropy, represented by the
constraint\- $H(p,\rho,\nu)$=$0$, is not only a direct path to
avoid the loss of the general relativity limit, but also a natural
way to reduce the degrees of freedom on the brane. This is an
enormous simplification which has proven to be useful in searching
physically relevant exact solution on braneworld. Furthermore,
this method works equally for analytic as well for numerical
methods. Hence the condition of minimal anisotropy might help in
the search of not only analytic physically acceptable, but also
numerical solutions when a non-uniform distribution is considered.
Thus the role played by the density gradients as a source of Weyl
stresses in the interior could be studied.

The work developed in this paper represents the point of view of a
brane observer. Hence the solution found here, a physically
acceptable one, does not ensure that the bulk eventually
constructed will not be plagued with singularities or any other
problem. However, since the condition of minimal anisotropy on the
brane ensures the correct limit at low energies, it could be used
when the bulk configuration is investigated, thus some general
features of the five dimensional bulk might be elucidated
\footnote{The condition H=0 ensures the minimal anisotropy and
therefore has the virtue of being less "expensive" for the
embedding+bulk so that it might be seriously consider as a
possible regular candidate for a brane star \cite{CristianoPC}.}.
This is currently been investigated.

\section*{Acknowledgments}

The author thanks Cristiano Germani for his valuable comments.




\end{document}